# Size-dependent piezoelectricity

Ali R. Hadjesfandiari

*Department of Mechanical and Aerospace Engineering*
*University at Buffalo, State University of New York*
*Buffalo, NY 14260 USA*

ah@buffalo.edu

June 28, 2012

**Abstract**

In this paper, a consistent theory is developed for size-dependent piezoelectricity in dielectric solids. This theory shows that electric polarization can be generated as the result of coupling to the mean curvature tensor, unlike previous flexoelectric theories that postulate such couplings with other forms of curvature and more general strain gradient terms ignoring the possible couple- stresses. The present formulation represents an extension of recent work that establishes a consistent size-dependent theory for solid mechanics. Here by including scale-dependent measures in the energy equation, the general expressions for force- and couple-stresses, as well as electric displacement, are obtained. Next, the constitutive relations, displacement formulations, the uniqueness theorem and the reciprocal theorem for the corresponding linear small deformation size-dependent piezoelectricity are developed. As with existing flexoelectric formulations, one finds that the piezoelectric effect can also exist in isotropic materials, although in the present theory the coupling is strictly through the skew-symmetric mean curvature tensor. In the last portion of the paper, this isotropic case is considered in detail by developing the corresponding boundary value problem for two dimensional analyses and obtaining a closed form solution for an isotropic dielectric cylinder.

## 1.  Introduction

Recent developments in micromechanics, nanomechanics and nanotechnology require advanced size dependent electromechanical modeling of coupled phenomena, such as piezoelectricity.

Classical piezoelectricity describes the relation between electric polarization and strain in non-centrosymmetric dielectrics at the macro-scale (Cady, 1964). However, some experiments have reported about size-effect phenomena of piezoelectric solids and linear electromechanical coupling in isotropic materials (Mishima, et al., 1997; Shvartsman et al., 2002; Buhlmann et al., 2002; Cross, 2006; Harden et al., 2006; Zhu et al., 2006; Baskaran et al., 2011; Catalan, et al., 2011). The classical theory cannot address this size dependency, because it considers that matter is continuously distributed throughout the body by neglecting its microstructure. Therefore, it is necessary to develop a size-dependent piezoelectricity, which accounts for the microstructure of the material by introducing higher gradient of deformation. Wang et al. (2004) have developed a size-dependent piezoelectric theory by considering the rotation gradient effect in the framework of the couple stress theory. In this formulation the electric polarization is related to the macroscopic rotation gradient. However, the theory suffers from its dependence on an underlying inconsistent couple stress theory. In some circles this size-dependent character for linear response is known as the flexoelectric effect (Kogan, 1964; Meyer, 1969), where the dielectric polarization is related to the macroscopic strain gradient or curvature strain. This theory predicts that in principle the flexoelectric effect is nonzero for all dielectrics, including the isotropic ones. Although there are some developments in this direction (Tagantsev, 1986; Maranganti et al., 2004; Sharma, 2007, Eliseev et al., 2009), these theories also suffer from the use of different inconsistent second order gradients of deformation, as well as ignoring the possible couple-stress effect. There have been some experimental studies, which correlate their data with these theories (e.g., Cross, 2006; Harden et al., 2006; Zhu et al., 2006; Baskaran et al., 2011; Catalan et al., 2011; Morozovska et al., 2012).

Thus, the first step toward developing consistent size-dependent electromechanical theories is the establishment of the consistent size-dependent continuum mechanics theory. Recently, Hadjesfandiari and Dargush (2011) have resolved the troubles in the existing size-dependent continuum mechanics. This progress shows that the couple-stress tensor has a vectorial character and that the body couple is not distinguishable from the body force. In this theory, the stresses are fully determinate and the measure of deformation is the mean curvature tensor, which is the skew-symmetrical part of the macroscopic rotation gradient. This development can be considered the completion of the works of Mindlin and Tiersten (1962), and Koiter (1964).

Furthermore, this size-dependent continuum mechanics must provide the fundamental base for developing different mechanical and electromechanical formulations that may govern the behavior of solid continua at the smallest scales. Here, the consistent size-dependent piezoelectric theory is developed, which shows that the size-dependent piezoelectric effect is related to the mean curvature tensor.

In the following section, we provide an overview of the electromechanical equations. This includes the equations for the kinematics, kinetics and quasi-electrostatics of size-dependent small deformation continuum mechanics. In Section 3, we consider the energy equation and its consequences based on the first law of thermodynamics for dielectric materials. In Section 4, the constitutive relations for linear elastic piezoelectric materials also are derived along with the governing equations. Next, we develop two weak formulations in Section 5, which are used to establish conditions for uniqueness and to derive the reciprocal identity. Section 6 provides the general theory for isotropic linear material and the details for two dimensional cases are derived, including the closed form solution for polarization of a long cylinder in a uniform electric field. Finally, Section 7 contains a summary and some general conclusions.

## 2. Basic size-dependent electromechanical equations

Let us take the three dimensional coordinate system $x_1 x_2 x_3$ as the reference frame with unit base vectors $\mathbf{e}_1$, $\mathbf{e}_2$ and $\mathbf{e}_3$. Consider a piezoelectric material continuum occupying a volume $V$ bounded by a surface $S$. In size-dependent continuum theory, the interaction in the body is represented by force-stress $\sigma_{ij}$ and couple-stress $\mu_{ij}$ tensors. The force-traction vector $t_i^{(n)}$ and moment-traction vector $m_i^{(n)}$ at a point on surface element $dS$ with unit normal vector $n_i$ are given by

$$t_i^{(n)} = \sigma_{ji} n_j \quad (1)$$

$$m_i^{(n)} = \mu_{ji} n_j \quad (2)$$

The force-stress tensor is generally non-symmetric and can be decomposed as

$$\sigma_{ji} = \sigma_{(ji)} + \sigma_{[ji]} \tag{3}$$

where $\sigma_{(ji)}$ and $\sigma_{[ji]}$ are the symmetric and skew-symmetric parts, respectively. Hadjesfandiari and Dargush (2011) have shown that, the couple-stress tensor is skew-symmetrical

$$\mu_{ji} = -\mu_{ij} \tag{4}$$

This means the moment-traction $m_i^{(n)}$ given by (2) is tangent to the surface. As a result, the couple-stress tensor $\mu_{ij}$ creates only bending moment-traction on any arbitrary surface in matter.

We can define the couple-stress vector $\mu_i$ dual to the tensor $\mu_{ij}$ as

$$\mu_i = \frac{1}{2}\varepsilon_{ijk}\mu_{kj} \tag{5}$$

where $\varepsilon_{ijk}$ is the permutation tensor or Levi-Civita symbol. This relation can also be written in the form

$$\varepsilon_{ijk}\mu_k = \mu_{ji} \tag{6}$$

Consequently, the surface moment-traction vector $m_i^{(n)}$ reduces to

$$m_i^{(n)} = \mu_{ji}n_j = \varepsilon_{ijk}n_j\mu_k \tag{7}$$

which is obviously tangent to the surface.

To formulate the fundamental equations, we consider an arbitrary part of this electromechanical body occupying a volume $V_a$ enclosed by boundary surface $S_a$. In infinitesimal deformation theory, the displacement vector field $\mathbf{u}(\mathbf{x},t)$ is so small that the velocity and acceleration fields can be approximated by $\dot{\mathbf{u}}$ and $\ddot{\mathbf{u}}$, respectively. As a result, the linear and angular equations of motion for this part of the body are written as

$$\int_{S_a} t_i^{(n)}dS + \int_{V_a} F_i dV = \int_{V_a} \rho \ddot{u}_i dV \tag{8}$$

$$\int_{S_a}\left[\varepsilon_{ijk}x_j t_k^{(n)} + m_i^{(n)}\right]dS + \int_{V_a}\varepsilon_{ijk}x_j F_k dV = \int_{V}\varepsilon_{ijk}x_j \rho \ddot{u}_k dV \tag{9}$$

where $F_i$ is the body force per unit volume of the body, and $\rho$ is the mass density. Hadjesfandiari and Dargush (2011) have shown that the body couple density is not distinguishable from body force in size-dependent couple stress continuum mechanics and its effect is simply equivalent to a system of body force and surface traction.

By using the relations (1) and (2) for tractions in the equations of motion (8) and (9), along with the divergence theorem, and noticing the arbitrariness of volume $V_a$, we finally obtain the differential form of the equations of motion as

$$\sigma_{ji,j} + F_i = \rho \ddot{u}_i \tag{10}$$

$$\mu_{ji,j} + \varepsilon_{ijk}\sigma_{jk} = 0 \tag{11}$$

Since the couple-stress tensor $\mu_{ji}$ is skew-symmetric, the angular equilibrium equation (11) gives the skew-symmetric part of the force-stress tensor as

$$\sigma_{[ji]} = -\frac{1}{2}\varepsilon_{ipq}\mu_{qp,j} = -\mu_{[i,j]} \tag{12}$$

In infinitesimal deformation theory, we may assume

$$\left|\frac{\partial u_i}{\partial x_j}\right| << 1 \quad , \quad \left|\frac{\partial^2 u_i}{\partial x_j \partial x_k}\right| << \frac{1}{l_S} \tag{13}$$

where $l_S$ is the smallest characteristic length in the body. Therefore, the infinitesimal strain, rotation and mean curvature tensors are defined as

$$e_{ij} = u_{(i,j)} = \frac{1}{2}\left(u_{i,j} + u_{j,i}\right) \tag{14}$$

$$\omega_{ij} = u_{[i,j]} = \frac{1}{2}\left(u_{i,j} - u_{j,i}\right) \tag{15}$$

$$\kappa_{ij} = \omega_{[i,j]} = \frac{1}{2}\left(\omega_{i,j} - \omega_{j,i}\right) \tag{16}$$

respectively. Since the tensors $\omega_{ij}$ and $\kappa_{ij}$ are skew-symmetrical, one can introduce their corresponding dual rotation and mean curvature vectors as

$$\omega_i = \frac{1}{2}\varepsilon_{ijk}\omega_{kj} \tag{17}$$

$$\kappa_i = \frac{1}{2}\varepsilon_{ijk}\kappa_{kj} \tag{18}$$

Consider the orthogonal planes parallel to $x_2 x_3$, $x_3 x_1$ and $x_1 x_2$ at each point before deformation. These planes under deformation transform to three surfaces with mean curvatures, having the components $\kappa_1$, $\kappa_2$ and $\kappa_3$ in the $x_1$, $x_2$ and $x_3$ directions, respectively. By inserting from (16) into (18), we obtain

$$\kappa_i = \frac{1}{4}u_{k,ki} - \frac{1}{4}\nabla^2 u_i \tag{19}$$

which also shows

$$\kappa_i = \frac{1}{2}\omega_{ji,j} \tag{20}$$

For a quasistatic electric field $\mathbf{E}$, in which the effect of induced magnetic field in the material is neglected, we have the electrostatic relation

$$\nabla \times \mathbf{E} = 0 \tag{21}$$

$$\varepsilon_{ijk} E_{k,j} = 0 \tag{22}$$

This simply shows that

$$E_{i,j} - E_{j,i} = 0 \tag{23}$$

Therefore, as is well-known in this case, the electric field $\mathbf{E}$ can be represented by the electric potential $\phi$, such that

$$\mathbf{E} = -\nabla \phi \tag{24}$$

or

$$E_i = -\phi_{,i} \tag{25}$$

The electric field and deformation can induce polarization $\mathbf{P}$ in the dielectric material. The electric displacement vector $\mathbf{D}$ is defined by

$$\mathbf{D} = \epsilon_0 \mathbf{E} + \mathbf{P} \tag{26}$$

$$D_i = \epsilon_0 E_i + P_i \tag{27}$$

where $\epsilon_0$ is the permittivity of free space. The normal electric displacement on the surface is defined by the scalar

$$d = D_i n_i \qquad (28)$$

The electric Gauss law for the volume $V_a$ is written as

$$\int_{S_a} d\, dS = \int_{V_a} \rho_E dV \qquad (29)$$

or

$$\int_{S_a} D_i n_i dS = \int_{V_a} \rho_E dV \qquad (30)$$

where $\rho_E$ is the electric charge density in the volume. By using the divergence theorem, and noticing the arbitrariness of volume $V_a$, we obtain the differential form of the Gauss law as

$$D_{i,i} = \rho_E \qquad (31)$$

It seems we should mention the character of the electric potential and normal electric displacement across boundary $S$ between two different materials carrying a free surface charge with intensity $q_S$. The electric potential is continuous across the interface, where

$$\lfloor \phi \rfloor = 0 \quad \text{on} \quad S \qquad (32)$$

with $\lfloor \ \rfloor$ denoting the jump across the interface. By applying Gauss law (29) to a thin Gaussian pillbox extending infinitesimally into the domains on either side of the boundary, we obtain

$$\lfloor d \rfloor = q_S \quad \text{on} \quad S \qquad (33)$$

Interestingly, this relation shows that if there is no free surface charge $q_S$, then $d$ is also continuous across the interface.

What has been presented so far is a continuum mechanics theory of electromechanical materials with quasistatic electric field, independent of the material properties. The governing electromechanical equations in the volume $V$ are

$$\sigma_{ji,j} + F_i = \rho \ddot{u}_i \qquad (34)$$

$$\sigma_{[ji]} + \mu_{[i,j]} = 0 \qquad (35)$$

$$D_{i,i} = \rho_E \qquad (36)$$

subject to some prescribed compatible boundary conditions on the boundary $S$. From a mathematical point of view, we can specify either the displacement vector $u_i$ or the force-traction vector $t_i^{(n)}$, the tangential component of the rotation vector $\omega_i$ or the tangential component of the moment-traction vector $m_i^{(n)}$, and the electric potential $\phi$ or the normal electric displacement $d$. However, in practice, the actual boundary is usually free of moment traction ($m_i^{(n)} = 0$) everywhere on $S$. Therefore, the tangential component of $\omega_i$, usually is not specified on the actual boundary $S$.

The equations (34)-(36) on their own are not enough to describe the electromechanical response of any particular material. To complete the specification, we need to define the electromechanical constitutive equations. For this we need to consider the energy equation.

## 3. Energy equation for piezoelectric material and constitutive relations

The energy equation for the electromechanical elastic medium in volume $V_a$, which undergoes small deformation and quasistatic polarization, is

$$\frac{\partial}{\partial t} \int_{V_a} \left( \frac{1}{2} \rho \dot{u}_i \dot{u}_i + U \right) dV = \int_{S_a} \left( t_i^{(n)} \dot{u}_i + m_i^{(n)} \dot{\omega}_i - \dot{d}\varphi \right) dS + \int_{V_a} \left( F_i \dot{u}_i + \varphi \dot{\rho}_E \right) dV \qquad (37)$$

where $U$ is the internal energy per unit volume. This equation shows that the rate of change of total energy of the system in volume $V_a$ is equivalent to the power of the external forces, moments and electric field.

By using the relations (34-36) along with the divergence theorem, and noticing the arbitrariness of volume $V_a$, one can obtain

$$\dot{U} = \sigma_{(ji)} \dot{e}_{ij} + \mu_{ji} \dot{\kappa}_{ij} - \varphi_{,i} \dot{D}_i \qquad (38)$$

This equation is the first law of thermodynanics for the size-dependent electromechanical elastic medium in differential form, which can also be written as

$$\dot{U} = \sigma_{(ji)}\dot{e}_{ij} + \mu_{ji}\dot{\kappa}_{ij} + E_i\dot{D}_i \qquad (39)$$

or

$$\dot{U} = \sigma_{(ji)}\dot{e}_{ij} - 2\mu_i\dot{\kappa}_i + E_i\dot{D}_i \qquad (40)$$

The relation (40) shows that for an elastic piezoelectric solid with couple stress effects, the internal energy $U$ depends not only on the strain tensor **e** and the electric displacement vector **D**, but also on the mean curvature tensor **κ**, that is

$$U = U(e_{ij}, \kappa_{ij}, D_i) \qquad (41)$$

$$U = U(e_{ij}, \kappa_i, D_i) \qquad (42)$$

By using the Legendre transformation, we define the specific electric enthalpy as

$$H = U - E_i D_i \qquad (43)$$

Then differentiating with respect to time, we obtain

$$\dot{H} = \dot{U} - \dot{E}_i D_i - E_i \dot{D}_i \qquad (44)$$

which with equation (40) for $\dot{U}$ yields

$$\dot{H} = \sigma_{(ji)}\dot{e}_{ij} + \mu_{ji}\dot{\kappa}_{ij} - D_i\dot{E}_i \qquad (45)$$

or

$$\dot{H} = \sigma_{(ji)}\dot{e}_{ij} - 2\mu_i\dot{\kappa}_i - D_i\dot{E}_i \qquad (46)$$

This equation implies that

$$H = H(e_{ij}, \kappa_{ij}, E_i) \qquad (47)$$

or

$$H = H(e_{ij}, \kappa_i, E_i) \qquad (48)$$

If we differentiate these forms of $H$ with respect to time, we obtain

$$\dot{H} = \frac{\partial H}{\partial e_{ij}}\dot{e}_{ij} + \frac{\partial H}{\partial \kappa_{ij}}\dot{\kappa}_{ij} + \frac{\partial H}{\partial E_i}\dot{E}_i \qquad (49)$$

or

$$\dot{H} = \frac{\partial H}{\partial e_{ij}} \dot{e}_{ij} + \frac{\partial H}{\partial \kappa_i} \dot{\kappa}_i + \frac{\partial H}{\partial E_i} \dot{E}_i \tag{50}$$

By comparing (49) with (45) and considering the arbitrariness of $\dot{e}_{ij}$, $\dot{\kappa}_{ij}$ and $\dot{E}_i$, we find the following constitutive relations for the symmetric part of force-stress tensor $\sigma_{(ji)}$, the couple-stress tensor $\mu_{ji}$ and the electric displacement vector $D_i$:

$$\sigma_{(ji)} = \frac{1}{2}\left(\frac{\partial H}{\partial e_{ij}} + \frac{\partial H}{\partial e_{ji}}\right) \tag{51}$$

$$\mu_{ji} = \frac{1}{2}\left(\frac{\partial H}{\partial \kappa_{ij}} - \frac{\partial H}{\partial \kappa_{ji}}\right) \tag{52}$$

$$D_i = -\frac{\partial H}{\partial E_i} \tag{53}$$

If we further agree to construct the functional $H$, such that

$$\frac{\partial H}{\partial e_{ij}} = \frac{\partial H}{\partial e_{ji}} \tag{54}$$

$$\frac{\partial H}{\partial \kappa_{ij}} = -\frac{\partial H}{\partial \kappa_{ji}} \tag{55}$$

we can write in place of (51) and (52)

$$\sigma_{(ji)} = \frac{\partial H}{\partial e_{ij}} \tag{56}$$

$$\mu_{ji} = \frac{\partial H}{\partial \kappa_{ij}} \tag{57}$$

It should be noticed that by comparing (46) and (50), one obtains

$$\mu_i = -\frac{1}{2}\frac{\partial H}{\partial \kappa_i} \tag{58}$$

for the couple-stress vector, which is more suitable in the following. By using this in the relation (12), we obtain the skew-symmetric part of the force-stress tensor as

$$\sigma_{[ji]} = -\mu_{[i,j]} = \frac{1}{4}\left(\frac{\partial H}{\partial \kappa_i}\right)_{,j} - \frac{1}{4}\left(\frac{\partial H}{\partial \kappa_j}\right)_{,i} \tag{59}$$

Therefore, for the total force-stresses, we have

$$\sigma_{ji} = \frac{\partial H}{\partial e_{ij}} + \frac{1}{4}\left(\frac{\partial H}{\partial \kappa_i}\right)_{,j} - \frac{1}{4}\left(\frac{\partial H}{\partial \kappa_j}\right)_{,i} \tag{60}$$

It is seen that the remaining electromechanical relations are the linear equation of motion and electric Gauss law

$$\sigma_{ji,j} + F_i = \rho \ddot{u}_i \tag{61}$$

$$D_{i,i} = \rho_E \tag{62}$$

**4. Linear piezoelectricity theory**

For linear elastic size-dependent piezoelectricity theory, we consider the homogeneous quadratic form for $H$

$$H = \frac{1}{2} A_{ijkl} e_{ij} e_{kl} + \frac{1}{2} B_{ijkl} \kappa_{ij} \kappa_{kl} + C_{ijkl} e_{ij} \kappa_{kl} - \frac{1}{2} \epsilon_{ij} E_i E_j - \alpha_{ijk} E_i e_{jk} - \beta_{ijk} E_i \kappa_{jk} \tag{63}$$

The first three terms represent the most general form of the elastic energy density. The tensors $A_{ijkl}$, $B_{ijkl}$ and $C_{ijkl}$ contain the elastic constitutive coefficients and are such that the elastic energy is positive definite. As a result, tensors $A_{ijkl}$ and $B_{ijkl}$ are positive definite. The tensor $A_{ijkl}$ is actually equivalent to its corresponding tensor in Cauchy elasticity. The tensor $\epsilon_{ij}$ is the permittivity or dielectric tensor, which is also positive definite. The tensors $\alpha_{ijk}$ and $\beta_{ijk}$ represent the piezoelectric character of the material. While the tensor $\alpha_{ijk}$ is the classical piezoelectric tensor, the tensor $\beta_{ijk}$ is the size-dependent coupling term between the electric field and the mean curvature tensor. The tensor $\beta_{ijk}$ may be called the flexoelectric tensor, which accounts for the microstructure of the material. The symmetry and skew-symmetry relations are

$$A_{ijkl} = A_{klij} = A_{jikl} \tag{64}$$

$$B_{ijkl} = B_{klij} = -B_{jikl} \tag{65}$$

$$C_{ijkl} = C_{jikl} = -C_{ijlk} \tag{66}$$

$$\epsilon_{ij} = \epsilon_{ji} \tag{67}$$

$$\alpha_{ijk} = \alpha_{ikj} \tag{68}$$

$$\beta_{ijk} = -\beta_{ikj} \tag{69}$$

For the most general case, the number of distinct components for $A_{ijkl}$, $B_{ijkl}$, $C_{ijkl}$, $\epsilon_{ij}$, $\alpha_{ijk}$ and $\beta_{ijk}$ are 21, 6, 18, 6, 18 and 9, respectively. Therefore, the most general linear elastic piezoelectric anisotropic material is described by 78 independent constitutive coefficients. It is interesting to note that the enthalpy density $H$ can also be written in terms of the mean curvature vector as

$$H = \frac{1}{2} A_{ijkl} e_{ij} e_{kl} + \frac{1}{2} B_{ij} \kappa_i \kappa_j + C_{ijk} e_{ij} \kappa_k - \frac{1}{2} \epsilon_{ij} E_i E_j - \alpha_{ijk} E_i e_{jk} - \beta_{ij} E_i \kappa_j \tag{70}$$

where

$$B_{ijkl} = \frac{1}{4} \varepsilon_{ijp} \varepsilon_{klq} B_{pq} \tag{71}$$

$$C_{ijkl} = \frac{1}{2} C_{ijm} \varepsilon_{mlk} \tag{72}$$

$$\beta_{ijk} = \frac{1}{2} \beta_{im} \varepsilon_{mkj} \tag{73}$$

and the symmetry relations

$$B_{ij} = B_{ji} \tag{74}$$

$$C_{ijk} = C_{jik} \tag{75}$$

hold. It is seen that the tensor $B_{ij}$ is positive definite, and there is no general symmetry condition for the tensor $\beta_{ij}$. Note that the number of distinct components for $B_{ij}$, $C_{ijk}$, and $\beta_{ij}$ are 6, 18 and 9, respectively. We should also notice that there is no restriction on the piezoelectric and flexoelectric tensors $\alpha_{ijk}$ and $\beta_{ij}$ for a well posed linear size-dependent piezoelectric boundary value problem.

By using the enthalpy density (70) in the general relations (56), (58) and (53), we obtain the following constitutive relations

$$\sigma_{(ji)} = A_{ijkl} e_{kl} + C_{ijk} \kappa_k - \alpha_{kij} E_k \tag{76}$$

$$\mu_i = -\frac{1}{2}B_{ij}\kappa_j - \frac{1}{2}C_{kji}e_{kj} + \frac{1}{2}\beta_{ji}E_j \tag{77}$$

$$D_i = \epsilon_{ij}E_j + \alpha_{ijk}e_{jk} + \beta_{ij}\kappa_j \tag{78}$$

The skew-symmetric part of the force-stress tensor is found as

$$\sigma_{[ji]} = -\mu_{[i,j]} = \frac{1}{4}B_{im}\kappa_{m,j} - \frac{1}{4}B_{jm}\kappa_{m,i} + \frac{1}{4}C_{kmi}e_{km,j} - \frac{1}{4}C_{kmj}e_{km,i} - \frac{1}{4}\beta_{mi}E_{m,j} + \frac{1}{4}\beta_{mj}E_{m,i} \tag{79}$$

Therefore, the constitutive relation for the total force-stress tensor is

$$\sigma_{ji} = A_{ijkl}e_{kl} + C_{ijk}\kappa_k - \alpha_{kji}E_k + \frac{1}{4}B_{im}\kappa_{m,j} - \frac{1}{4}B_{jm}\kappa_{m,i}$$
$$+ \frac{1}{4}C_{kmi}e_{km,j} - \frac{1}{4}C_{kmj}e_{km,i} - \frac{1}{4}\beta_{mi}E_{m,j} + \frac{1}{4}\beta_{mj}E_{m,i} \tag{80}$$

For the polarization vector, we have the relation

$$P_i = D_i - \epsilon_0 E_i = \left(\epsilon_{ij} - \epsilon_0 \delta_{ij}\right)E_j + \alpha_{ijk}e_{jk} + \beta_{ij}\kappa_j \tag{81}$$

Interestingly, for the internal energy density function $U$, we obtain

$$U = H + E_i D_i = \frac{1}{2}A_{ijkl}e_{ij}e_{kl} + \frac{1}{2}B_{ij}\kappa_i\kappa_j + C_{ijk}e_{ij}\kappa_k + \frac{1}{2}\epsilon_{ij}E_i E_j \tag{82}$$

which is a positive definite quadratic form without explicit piezoelectric and flexoelectric coupling.

When the constitutive relations force-stress tensor (80) and electric displacement vector (78) are written in terms of displacements and electric potential, we obtain

$$\sigma_{ji} = A_{ijkl}u_{k,l} + \frac{1}{4}C_{ijk}\left(u_{m,mk} - \nabla^2 u_k\right) + \frac{1}{4}C_{kmi}u_{k,mj} - \frac{1}{4}C_{kmj}u_{k,mi}$$
$$+ \frac{1}{16}B_{ik}\left(u_{m,mkj} - \nabla^2 u_{k,j}\right) - \frac{1}{16}B_{jk}\left(u_{m,mki} - \nabla^2 u_{k,i}\right) + \alpha_{kji}\phi_{,k} + \frac{1}{4}\beta_{mi}\phi_{,mj} - \frac{1}{4}\beta_{mj}\phi_{,mi} \tag{83}$$

$$D_i = -\epsilon_{ij}\phi_{,j} + \alpha_{ijk}u_{j,k} + \frac{1}{4}\beta_{ij}\left(u_{m,mj} - \nabla^2 u_j\right) \tag{84}$$

By carrying these forms into the linear equation of motion (61) and electric Gauss law (62), one obtains the governing equations for size-dependent piezoelectricity as follows:

$$A_{ijkl}u_{k,lj} + \frac{1}{4}C_{ijk}\left(u_{m,mjk} - \nabla^2 u_{k,j}\right) + \frac{1}{4}C_{kmi}\nabla^2 u_{k,m} - \frac{1}{4}C_{kmj}u_{k,mij}$$
$$+ \frac{1}{16}B_{ik}\left(\nabla^2 u_{m,mk} - \nabla^2\nabla^2 u_k\right) - \frac{1}{16}B_{jk}\left(u_{m,mkij} - \nabla^2 u_{k,ij}\right) \quad (85)$$
$$+ \alpha_{kji}\phi_{,kj} + \frac{1}{4}\beta_{mi}\nabla^2\phi_{,m} - \frac{1}{4}\beta_{mj}\phi_{,mij} + F_i = \rho\ddot{u}_i$$

$$\epsilon_{ij}\phi_{,ij} - \alpha_{ijk}u_{j,ik} - \frac{1}{4}\beta_{ij}\left(u_{m,mij} - \nabla^2 u_{j,i}\right) + \rho_E = 0 \quad (86)$$

It is obvious that the displacement vector and electric potential equations are coupled in these equations. As we mentioned, the prescribed boundary conditions on the surface of the body can be any compatible combination of $u_i, \omega_i, \phi$ and $t_i^{(n)}, m_i^{(n)}, d$. The force-traction vector $t_i^{(n)}$, the moment-traction $m_i^{(n)}$ and normal electric displacement $d$ at any point on a surface $S$, with outward unit normal vector $n_i$, are

$$t_i^{(n)} = \sigma_{ji}n_j$$
$$= \begin{pmatrix} A_{ijkl}e_{kl} + C_{ijk}\kappa_k - \alpha_{kji}E_k + \frac{1}{4}B_{im}\kappa_{m,j} - \frac{1}{4}B_{jm}\kappa_{m,i} \\ + \frac{1}{4}C_{kmi}e_{km,j} - \frac{1}{4}C_{kmj}e_{km,i} - \frac{1}{4}\beta_{mi}E_{m,j} + \frac{1}{4}\beta_{mj}E_{m,i} \end{pmatrix} n_j \quad (87)$$

$$m_i^{(n)} = \varepsilon_{ijk}\mu_k n_j = \varepsilon_{ijk}\left(-\frac{1}{2}B_{km}\kappa_m - \frac{1}{2}C_{mnk}e_{mn} + \frac{1}{2}\beta_{mk}E_m\right)n_j \quad (88)$$

$$d = D_i n_i = \left(\epsilon_{ij}E_j + \alpha_{ijk}e_{jk} + \beta_{ij}\kappa_j\right)n_i \quad (89)$$

It should be noticed that the flexoelectric effect always appears along with couple-stresses. This means that if we neglect the couple stresses ($B_{ij} = 0$, $C_{ijk} = 0$), all other size-dependent effects such as flexoelectricity must be neglected as well ($\beta_{ij} = 0$).

## 5. Weak formulations and their consequences

Weak formulations or virtual work theorems have many applications in all aspects of continuum mechanics, such as variational and integral equation methods. Weak formulations can also be used in exploring conditions of uniqueness. Therefore, we derive two forms of these principles for the static state of size-dependent piezoelectricity as follows.

### 5.1. Weak forms of equilibrium equations

Consider the part of the body occupying a fixed volume $V$ bounded by boundary surface $S$. The standard form of the equilibrium equations for this electromechanical medium in the static case are given by

$$\sigma_{ji,j} + F_i = 0 \tag{90}$$

$$\mu_{ji,j} + \varepsilon_{ijk}\sigma_{jk} = 0 \tag{91}$$

$$D_{i,i} = \rho_E \tag{92}$$

Suppose arbitrary differentiable displacement variation $\delta u_i$ and electric potential variation $\delta\phi$ in the domain, where their corresponding angular rotation and electric fields are

$$\delta\omega_i = \frac{1}{2}\varepsilon_{ijk}\delta u_{k,j} \tag{93}$$

$$\delta E_i = -\delta\phi_{,i} \tag{94}$$

Let us multiply (91) by the virtual displacement $\delta u_i$ and integrate over the volume, and also multiply (92) by its corresponding virtual angular rotation $\delta\omega_i$ and integrate over the volume as well. Therefore, we have

$$\int_V (\sigma_{ji,j} + F_i)\delta u_i \, dV = 0 \tag{95}$$

$$\int_V (\mu_{ji,j} + \varepsilon_{ijk}\sigma_{jk})\delta\omega_i \, dV = 0 \tag{96}$$

By noticing the relation

$$\sigma_{ji,j}\delta u_i = (\sigma_{ji}\delta u_i)_{,j} - \sigma_{ji}\delta u_{i,j} \tag{97}$$

and using the divergence theorem, the relation (95) becomes

$$\int_V \sigma_{ji} \delta u_{i,j} dV = \int_S t_i^{(n)} \delta u_i dS + \int_V F_i \delta u_i dV \tag{98}$$

Similarly, by using the relation

$$\mu_{ji,j} \delta \omega_i + \varepsilon_{ijk} \sigma_{jk} \delta \omega_i = \left( \mu_{ji} \delta \omega_i \right)_{,j} - \mu_{ji} \delta \omega_{i,j} - \sigma_{jk} \delta \omega_{jk} \tag{99}$$

equation (96) becomes

$$\int_V \mu_{ji} \delta \omega_{i,j} dV - \int_V \sigma_{ji} \delta \omega_{ij} dV = \int_S m_i^{(n)} \delta \omega_i dS \tag{100}$$

Then, by adding (98) and (100), we obtain

$$\int_V \mu_{ji} \delta \omega_{i,j} dV + \int_V \sigma_{ji} \left( \delta u_{i,j} - \delta \omega_{ij} \right) dV = \int_S t_i^{(n)} \delta u_i dS + \int_V F_i \delta u_i dV + \int_S m_i^{(n)} \delta \omega_i dS \tag{101}$$

However, by noticing the relation

$$\delta e_{ij} = \delta u_{i,j} - \delta \omega_{ij} \tag{102}$$

for compatible virtual displacement, we obtain the virtual work theorem as

$$\int_V \sigma_{ji} \delta e_{ij} dV + \int_V \mu_{ji} \delta \omega_{i,j} dV = \int_S t_i^{(n)} \delta u_i dS + \int_S m_i^{(n)} \delta \omega_i dS + \int_V F_i \delta u_i dV \tag{103}$$

Since $\delta e_{ij}$ is symmetric and $\mu_{ji}$ is skew-symmetric, we have

$$\sigma_{ji} \delta e_{ij} = \sigma_{(ji)} \delta e_{ij} \tag{104}$$

$$\mu_{ji} \delta \omega_{i,j} = \mu_{ji} \delta \kappa_{ij} \tag{105}$$

Thus, the principle of virtual work can be written as

$$\int_V \sigma_{(ji)} \delta e_{ij} dV + \int_V \mu_{ji} \delta \kappa_{ij} dV = \int_S t_i^{(n)} \delta u_i dS + \int_S m_i^{(n)} \delta \omega_i dS + \int_V F_i \delta u_i dV \tag{106}$$

or

$$\int_V \left( \sigma_{ji} \delta e_{ij} - 2 \mu_i \delta \kappa_i \right) dV = \int_S t_i^{(n)} \delta u_i dS + \int_S m_i^{(n)} \delta \omega_i dS + \int_V F_i \delta u_i dV \tag{107}$$

Now, let us multiply equation (92) by the virtual potential $\delta \phi$ and integrate over the volume

$$\int_V \left( D_{i,i} - \rho_E \right) \delta \phi dV = 0 \tag{108}$$

By noticing the relation

$$D_{i,i}\delta\phi - \rho_E\delta\phi = (D_i\delta\phi)_{,i} - D_i\delta\phi_{,i} - \rho_E\delta\phi = 0 \tag{109}$$

we obtain

$$D_i\delta E_i = -(D_i\delta\phi)_{,i} + \rho_E\delta\phi \tag{110}$$

Therefore, the relation (108) by using the divergence theorem, becomes

$$\int_V D_i\delta E_i dV = -\int_S d\delta\phi dS + \int_V \rho_E\delta\phi dV \tag{111}$$

which is the electrical analog of the virtual work theorem (106). Although these two virtual forms are independent, it turns out their combination is more useful in further investigation. By subtracting (106) and (111), we obtain the weak form

$$\int_V \left(\sigma_{ji}\delta e_{ij} + \mu_{ji}\delta\kappa_{ij} - D_i\delta E_i\right)dV$$
$$= \int_S t_i^{(n)}\delta u_i dS + \int_S m_i^{(n)}\delta\omega_i dS + \int_V F_i\delta u_i dV + \int_S d\delta\phi dS - \int_V \rho_E\delta\phi dV \tag{112}$$

We are also interested in a weak form corresponding to the energy equation (37). We can obtain this form analogously as follows. We consider the variation of electric displacement while holding the electric potential constant. Let $D_i$ be the actual electric displacement vector, which satisfies the Gauss law and boundary conditions. Now consider a virtual electric displacement vector $\delta D_i$ that satisfies the Gauss law

$$\delta D_{i,i} = \delta\rho_E \tag{113}$$

with virtual electric charge density $\delta\rho_E$ and boundary condition $\delta d = \delta D_i n_i$ on $S$. We multiply equation (113) by the potential $\phi$ and integrate over the volume

$$\int_V \left(\delta D_{i,i} - \delta\rho_E\right)\phi dV = 0 \tag{114}$$

Also by noticing the relation

$$\delta D_{i,i}\phi - \delta\rho_E\phi = (\delta D_i\phi)_{,i} - \delta D_i\phi_{,i} - \delta\rho_E\phi = 0 \tag{115}$$

we obtain

$$E_i\delta D_i = -(\delta D_i\phi)_{,i} + \delta\rho_E\phi \tag{116}$$

Therefore, we transform the relation (114) by using the divergence theorem to

$$\int_V E_i \delta D_i dV = -\int_S \phi \delta d\, dS + \int_V \phi \delta \rho_E dV \tag{117}$$

By adding the virtual theorems (106) and (117), we obtain

$$\begin{aligned}\int_V \left(\sigma_{ji}\delta e_{ij} + \mu_{ji}\delta \kappa_{ij} + E_i \delta D_i\right) dV \\ = \int_S t_i^{(n)} \delta u_i dS + \int_S m_i^{(n)} \delta \omega_i dS + \int_V F_i \delta u_i dV - \int_S \phi \delta d\, dS + \int_V \phi \delta \rho_E dV\end{aligned} \tag{118}$$

It should be noticed that alternative weak formulations, such as complementary virtual work, can also be developed. However, in this paper, we only consider the weak forms (112) and (118), which are used in the following sections.

### 5.2. Extremum potentials for piezoelectric material

For an elastic piezoelectric material, the weak form (112) reduces to

$$\int_V \delta H dV = \int_S t_i^{(n)} \delta u_i dS + \int_S m_i^{(n)} \delta \omega_i dS + \int_V F_i \delta u_i dV + \int_S d\delta\phi dS - \int_V \rho_E \delta\phi dV \tag{119}$$

By considering the compatible variations on the boundaries, we can write this as

$$\delta\left\{\int_V (H - F_i u_i + \rho_E \phi) dV - \int_{S_t} t_i^{(n)} u_i dS - \int_{S_m} m_i^{(n)} \omega_i dS - \int_{S_d} d\phi dS\right\} = 0 \tag{120}$$

where $S_t$, $S_m$ and $S_d$ are portion of surface at which $t_i^{(n)}$, $m_i^{(n)}$ and $d$ are prescribed, respectively. Therefore, by defining

$$\Pi_H = \int_V (H - F_i u_i + \rho_E \phi) dV - \int_{S_t} t_i^{(n)} u_i dS - \int_{S_m} m_i^{(n)} \omega_i dS - \int_{S_d} d\phi dS \tag{121}$$

we realize that the relation (120) shows that

$$\delta\Pi_H = 0 \tag{122}$$

The quantity $\Pi_H$ can be considered as the total electric enthalpy based electromechanical potential of the system. The relation (122) shows that the displacement and the electric potential fields satisfying equilibrium equations and boundary conditions must extremize $\Pi_H$.

We can obtain a second extremum for this elastic piezoelectric material by noticing that the weak form (118) reduces to

$$\int_V \delta U dV = \int_S t_i^{(n)} \delta u_i dS + \int_S m_i^{(n)} \delta \omega_i dS + \int_V F_i \delta u_i dV - \int_S \phi \delta d\, dS + \int_V \phi \delta \rho_E dV \tag{123}$$

Again by considering the compatible variations on the boundaries, we can write this as

$$\delta\left\{\int_V (U - F_i u_i - \phi \rho_E) dV - \int_{S_t} t_i^{(n)} u_i dS - \int_{S_m} m_i^{(n)} \omega_i dS + \int_{S_\phi} \phi d\, dS \right\} \quad (124)$$

where $S_t$, $S_m$ and $S_\phi$ are portion of surface at which $t_i^{(n)}$, $m_i^{(n)}$ and $\phi$ are prescribed, respectively. Let us define the total electromechanical energy potential of the system as

$$\Pi_U = \int_V (U - F_i u_i - \phi \rho_E) dV - \int_{S_t} t_i^{(n)} u_i dS - \int_{S_m} m_i^{(n)} \omega_i dS + \int_{S_\phi} \phi d\, dS \quad (125)$$

As a result, the relation (124) reduces to

$$\delta \Pi_U = 0 \quad (126)$$

which shows that the displacement and the electric charge density fields satisfying equilibrium equations and boundary conditions must extremize $\Pi_U$.

For linear size-dependent piezoelectricity, we also have the following interesting result. By replacing the virtual variations with the actual variations in the weak form (118), we obtain

$$\int_V \left(\sigma_{(ji)} e_{ij} - 2\mu_i \kappa_i + E_i D_i \right) dV$$
$$= \int_S t_i^{(n)} u_i dS + \int_S m_i^{(n)} \omega_i dS + \int_V F_i u_i dV - \int_S \phi d\, dS + \int_V \phi \rho_E dV \quad (127)$$

After using the constitutive relations (76)-(78) in the left hand side of this equation, we obtain

$$2\int_V U dV = \int_S t_i^{(n)} u_i dS + \int_S m_i^{(n)} \omega_i dS + \int_V F_i u_i dV - \int_S d\phi dS + \int_V \rho_E \phi dV \quad (128)$$

which gives twice the total internal energy in terms of the work of external body forces, surface tractions and electric displacement, and electric charges.

## 5.3. Uniqueness theorem for linear boundary value problems

Now we investigate the uniqueness of the corresponding linear size-dependent piezoelectric boundary value problem. The proof follows from the concept of electromechanical energy, similar to the approach for Cauchy elasticity.

Consider the general boundary value problem. The prescribed boundary conditions on the surface of the body can be any well-posed combination of vectors $u_i$ and $\omega_i$, $t_i^{(n)}$ and $m_i^{(n)}$, $\varphi$ and $d$ as discussed in Section 4. Assume that there exist two different solutions $\left\{ u_i^{(1)}, \phi^{(1)}, e_{ij}^{(1)}, \kappa_i^{(1)}, E_i^{(1)}, \sigma_{ji}^{(1)}, \mu_i^{(1)}, D_i^{(1)} \right\}$ and $\left\{ u_i^{(2)}, \phi^{(2)}, e_{ij}^{(2)}, \kappa_i^{(2)}, E_i^{(2)}, \sigma_{ji}^{(2)}, \mu_i^{(2)}, D_i^{(2)} \right\}$ to the same problem with identical body forces and boundary conditions. Thus, we have the equilibrium equations and electric Gauss law

$$\sigma_{ji,j}^{(\alpha)} + F_i = 0 \tag{129}$$

$$\sigma_{[ji]}^{(\alpha)} = -\mu_{[i,j]}^{(\alpha)} \tag{130}$$

$$D_{i,i}^{(\alpha)} = \rho_E \tag{131}$$

where

$$\sigma_{(ji)}^{(\alpha)} = A_{ijkl} e_{kl}^{(\alpha)} + C_{ijk} \kappa_k^{(\alpha)} - \alpha_{kij} E_k^{(\alpha)} \tag{132}$$

$$\mu_i^{(\alpha)} = -\frac{1}{2} B_{ij} \kappa_j^{(\alpha)} - \frac{1}{2} C_{kji} e_{kj}^{(\alpha)} + \frac{1}{2} \beta_{ji} E_j^{(\alpha)} \tag{133}$$

$$D_i^{(\alpha)} = \epsilon_{ij} E_j^{(\alpha)} + \alpha_{ijk} e_{jk}^{(\alpha)} + \beta_{ij} \kappa_j^{(\alpha)} \tag{134}$$

and the superscript $^{(\alpha)}$ references the solutions $^{(1)}$ and $^{(2)}$.

Let us now define the *difference solution* $\left\{ u_i', \phi', e_{ij}', \kappa_i', E_i', \sigma_{ji}', \mu_i', D_i' \right\}$

$$u_i' = u_i^{(2)} - u_i^{(1)} \tag{135a}$$

$$\phi' = \phi^{(2)} - \phi^{(1)} \tag{135b}$$

$$e_{ij}' = e_{ij}^{(2)} - e_{ij}^{(1)} \tag{135c}$$

$$\kappa_i' = \kappa_i^{(2)} - \kappa_i^{(1)} \tag{135d}$$

$$E_i' = E_i^{(2)} - E_i^{(1)} \tag{135e}$$

$$\sigma_{ji}' = \sigma_{ji}^{(2)} - \sigma_{ji}^{(1)} \tag{135f}$$

$$\mu_i' = \mu_i^{(2)} - \mu_i^{(1)} \tag{135g}$$

$$D_i' = D_i^{(2)} - D_i^{(1)} \tag{135h}$$

Since the solutions $\{u_i^{(1)}, \phi^{(1)}, e_{ij}^{(1)}, \kappa_i^{(1)}, E_i^{(1)}, \sigma_{ji}^{(1)}, \mu_i^{(1)}, D_i^{(1)}\}$ and $\{u_i^{(2)}, \phi^{(2)}, e_{ij}^{(2)}, \kappa_i^{(2)}, E_i^{(2)}, \sigma_{ji}^{(2)}, \mu_i^{(2)}, D_i^{(2)}\}$ correspond to the same body forces, electric charge densities and boundary conditions, the *difference solution* must satisfy the equilibrium equations

$$\sigma'_{ji,j} = 0 \tag{136}$$

$$\sigma'_{[ji]} = -\mu'_{[i,j]} \tag{137}$$

$$D'_{i,i} = 0 \tag{138}$$

with zero corresponding boundary conditions. Consequently, twice the total electromechanical energy (128) for the *difference solution* is

$$2\int_V U' dV = \int_V \left( A_{ijkl} e'_{ij} e'_{kl} + B_{ij} \kappa'_i \kappa'_j + 2C_{ijk} e'_{ij} \kappa'_k + \epsilon_{ij} E'_i E'_j \right) dV = 0 \tag{139}$$

Since the energy density of the *difference solution* $U'$ is non-negative, this relation requires

$$2U' = A_{ijkl} e'_{ij} e'_{kl} + B_{ij} \kappa'_i \kappa'_j + 2C_{ijk} e'_{ij} \kappa'_k + \epsilon_{ij} E'_i E'_j = 0 \quad \text{in } V \tag{140}$$

However, the tensors $A_{ijkl}$, $B_{ij}$ and $\epsilon_{ij}$ are positive definite and the tensor $C_{ijk}$ is such that the energy $U'$ is non-negative. Therefore the strain, curvature, electric field, and associated stresses and electric displacements for the *difference solution* must vanish

$$e'_{ij} = 0, \quad \kappa'_i = 0, \quad E'_i = 0, \quad \sigma'_{ij} = 0, \quad \mu'_i = 0, \quad D'_i = 0 \tag{141a-e}$$

These require that the *difference displacement* $u'_i$ and the *difference electric potential* $\phi'$ can be at most a rigid body motion and a constant potential, respectively. However, if displacement and potential are specified on parts of the boundary, then the *difference displacement and electric potential* vanish everywhere and we have

$$u_i^{(1)} = u_i^{(2)} \tag{142a}$$

$$\phi^{(1)} = \phi^{(2)} \tag{142b}$$

$$e_{ij}^{(1)} = e_{ij}^{(2)} \tag{142c}$$

$$\kappa_i^{(1)} = \kappa_i^{(2)} \tag{142d}$$

$$E_i^{(1)} = E_i^{(2)} \tag{142e}$$

$$\sigma_{ji}^{(1)} = \sigma_{ji}^{(2)} \tag{142f}$$

$$\mu_i^{(1)} = \mu_i^{(2)} \tag{142g}$$

$$D_i^{(1)} = D_i^{(2)} \tag{142h}$$

Therefore, the solution to the boundary value problem is unique. On the other hand, if only force- and moment- tractions are specified over the entire boundary, then the displacement is not unique and is determined only up to an arbitrary rigid body motion. Similarly, if only normal electric displacement is specified over the entire boundary, then the electric potential is not unique and is determined only up to an arbitrary constant.

## 5.4. Reciprocal theorem

We derive now the general reciprocal theorem for the equilibrium states of a linear elastic size-dependent piezoelectric material under different applied loads. Consider two sets of equilibrium states of compatible size-dependent piezoelectric solutions $\{u_i^{(1)}, \omega_i^{(1)}, \phi^{(1)}, t_i^{(n)(1)}, m_i^{(n)(1)}, d^{(1)}, F_i^{(1)}\}$ and $\{u_i^{(2)}, \omega_i^{(2)}, \phi^{(2)}, t_i^{(n)(2)}, m_i^{(n)(2)}, d^{(2)}, F_i^{(2)}\}$. Let us apply the weak form (112) in the forms

$$\int_V \left[ \sigma_{ji}^{(1)} e_{ij}^{(2)} - 2\mu_i^{(1)} \kappa_i^{(2)} - D_i^{(1)} E_i^{(2)} \right] dV$$
$$= \int_S t_i^{(n)(1)} u_i^{(2)} dS + \int_S m_i^{(n)(1)} \omega_i^{(2)} dS + \int_S d^{(1)} \phi^{(2)} dS + \int_V F_i^{(1)} u_i^{(2)} dV - \int_V \rho_E^{(1)} \phi^{(2)} dV \tag{143}$$

and

$$\int_V \left[ \sigma_{ji}^{(2)} e_{ij}^{(1)} - 2\mu_i^{(2)} \kappa_i^{(1)} - D_i^{(2)} E_i^{(1)} \right] dV$$
$$= \int_S t_i^{(n)(2)} u_i^{(1)} dS + \int_S m_i^{(n)(2)} \omega_i^{(1)} dS + \int_S d^{(2)} \phi^{(1)} dS + \int_V F_i^{(2)} u_i^{(1)} dV - \int_V \rho_E^{(2)} \phi^{(1)} dV \tag{144}$$

Using the constitutive relations (76)-(78), we obtain

$$\sigma_{ji}^{(1)} e_{ij}^{(2)} - 2\mu_i^{(1)} \kappa_i^{(2)} - D_i^{(1)} E_i^{(2)} = A_{ijkl} e_{kl}^{(1)} e_{ij}^{(2)} + C_{ijk} \kappa_k^{(1)} e_{ij}^{(2)} - \alpha_{kij} E_k^{(1)} e_{ij}^{(2)}$$
$$+ B_{ij} \kappa_j^{(1)} \kappa_i^{(2)} + C_{kji} e_{kj}^{(1)} \kappa_i^{(2)} - \beta_{ji} E_j^{(1)} \kappa_i^{(2)} \tag{145}$$
$$- \varepsilon_{ij} E_j^{(1)} E_i^{(2)} - \alpha_{ijk} e_{jk}^{(1)} E_i^{(2)} - \beta_{ij} \kappa_j^{(1)} E_i^{(2)}$$

$$\sigma_{ji}^{(2)} e_{ij}^{(1)} - 2\mu_i^{(2)} \kappa_i^{(1)} - D_i^{(2)} E_i^{(1)} = A_{ijkl} e_{kl}^{(2)} e_{ij}^{(1)} + C_{ijk} \kappa_k^{(2)} e_{ij}^{(1)} - \alpha_{kij} E_k^{(2)} e_{ij}^{(1)}$$
$$+ B_{ij} \kappa_j^{(2)} \kappa_i^{(1)} + C_{kji} e_{kj}^{(2)} \kappa_i^{(1)} - \beta_{ji} E_j^{(2)} \kappa_i^{(1)} \tag{146}$$
$$- \varepsilon_{ij} E_j^{(2)} E_i^{(1)} - \alpha_{ijk} e_{jk}^{(2)} E_i^{(1)} - \beta_{ij} \kappa_j^{(2)} E_i^{(1)}$$

Symmetry relations (64)-(69) shows that the left hand of (145) and (146) are the same and

$$\sigma_{ji}^{(1)}e_{ij}^{(2)} - 2\mu_i^{(1)}\kappa_i^{(2)} - D_i^{(1)}E_i^{(2)} = \sigma_{ji}^{(2)}e_{ij}^{(1)} - 2\mu_i^{(2)}\kappa_i^{(1)} - D_i^{(2)}E_i^{(1)} \tag{147}$$

Therefore, by comparing (143) and (144), we obtain the general reciprocal theorem for the linear size-dependent piezoelectric material as

$$\int_S t_i^{(n)(1)}u_i^{(2)}dS + \int_S m_i^{(n)(1)}\omega_i^{(2)}dS + \int_S \mathscr{d}^{(1)}\phi^{(2)}dS + \int_V F_i^{(1)}u_i^{(2)}dV - \int_V \rho_E^{(1)}\phi^{(2)}dV = \\ \int_S t_i^{(n)(2)}u_i^{(1)}dS + \int_S m_i^{(n)(2)}\omega_i^{(1)}dS + \int_S \mathscr{d}^{(2)}\phi^{(1)}dS + \int_V F_i^{(2)}u_i^{(1)}dV - \int_V \rho_E^{(2)}\phi^{(1)}dV \tag{148}$$

When there is no body force and charge density, this theorem reduces to

$$\int_S t_i^{(n)(1)}u_i^{(2)}dS + \int_S m_i^{(n)(1)}\omega_i^{(2)}dS + \int_S \mathscr{d}^{(1)}\phi^{(2)}dS = \int_S t_i^{(n)(2)}u_i^{(1)}dS + \int_S m_i^{(n)(2)}\omega_i^{(1)}dS + \int_S \mathscr{d}^{(2)}\phi^{(1)}dS \tag{149}$$

## 6. Isotropic linear piezoelectric material

### 6.1. General governing equations

The piezoelectric effect can exist in isotropic couple stress materials as we shall see. For an isotropic material, the symmetry relations require

$$A_{ijkl} = \lambda \delta_{ij}\delta_{kl} + \mu \delta_{ik}\delta_{jl} + \mu \delta_{il}\delta_{jk} \tag{150}$$

$$B_{ij} = 16\eta \delta_{ij} \tag{151}$$

$$C_{ijk} = 0 \tag{152}$$

$$\epsilon_{ij} = \epsilon \delta_{ij} \tag{153}$$

$$\alpha_{ijk} = 0 \tag{154}$$

$$\beta_{ij} = 4f\delta_{ij} \tag{155}$$

The moduli $\lambda$ and $\mu$ have the same meaning as the Lamé constants for an isotropic material in Cauchy elasticity. These two constants are related by

$$\lambda = 2\mu \frac{\nu}{1-2\nu} \tag{156}$$

where $\nu$ is the Poisson ratio. The material constants $\eta$ and $f$ account for couple stress and piezoelectricity effects in an isotropic material. The constant $f$ may be called the flexoelectric coefficient of the material.

As a result, the electromechanical enthalpy and energy densities become

$$H = \frac{1}{2}\lambda e_{jj}e_{kk} + \mu e_{ij}e_{ij} + 8\eta \kappa_i \kappa_i - \frac{1}{2}\epsilon E_i E_i - 4fE_i\kappa_i \qquad (157)$$

$$U = \frac{1}{2}\lambda e_{jj}e_{kk} + \mu e_{ij}e_{ij} + 8\eta \kappa_i \kappa_i + \frac{1}{2}\epsilon E_i E_i \qquad (158)$$

respectively. The following restrictions are necessary for positive definite energy density $U$

$$3\lambda + 2\mu > 0, \quad \mu > 0, \quad \eta > 0 \quad \epsilon > 0 \qquad (159)$$

We should notice that there is generally no restriction on the flexoelectric coefficient $f$. The ratio

$$\frac{\eta}{\mu} = l^2 \qquad (160)$$

specifies the characteristic material length $l$, which accounts for size-dependency in the small deformation couple stress elasticity theory under consideration here.

Then, the constitutive relations for the symmetric part of the force-stress tensor, the couple-stress vector, and the electric displacement vector can be written as

$$\sigma_{(ji)} = \lambda e_{kk}\delta_{ij} + 2\mu e_{ij} \qquad (161)$$

$$\mu_i = -8\mu l^2 \kappa_i + 2fE_i \qquad (162)$$

$$D_i = \epsilon E_i + 4f\kappa_i \qquad (163)$$

For the skew-symmetric part of the force-stress tensor, we have

$$\sigma_{[ji]} = -\mu_{[i,j]} = 2\mu l^2 \nabla^2 \omega_{ji} - f\left(E_{i,j} - E_{j,i}\right) \qquad (164)$$

The irrotational character of the electric field given by relation (23) shows that the last term disappears and

$$\sigma_{[ji]} = 2\mu l^2 \nabla^2 \omega_{ji} \tag{165}$$

or

$$\sigma_{[ji]} = 2\mu l^2 \varepsilon_{ijk} \nabla^2 \omega_k \tag{166}$$

Therefore, for the total force-stress tensor, we have

$$\sigma_{ji} = \lambda e_{kk} \delta_{ij} + 2\mu e_{ij} + 2\mu l^2 \nabla^2 \omega_{ji} \tag{167}$$

It should be noticed that the coefficient $f$ appears only in the couple-stress vector $\mu_i$ and polarization vector $D_i$, but not in the force-stress tensor $\sigma_{ji}$. As we mentioned before, if the effect of couple-stress is negligible ($l = 0$), the effect of flexoelectricity must be excluded ($f = 0$).

For the governing equations, we have

$$\left[\lambda + \mu\left(1 + l^2 \nabla^2\right)\right] u_{k,ki} + \mu\left(1 - l^2 \nabla^2\right) \nabla^2 u_i + F_i = \rho \ddot{u}_i \tag{168}$$

$$\epsilon \nabla^2 \phi + \rho_E = 0 \tag{169}$$

which are explicitly independent of $f$. However, the piezoelectric effect can exist due to the moment-traction

$$m_i^{(n)} = \varepsilon_{ijk} n_j \mu_k = \varepsilon_{ijk} n_j \left(-8\mu l^2 \kappa_k + 2f E_k\right) \tag{170}$$

which couples $\kappa_i$ and $E_i$ on the boundary. Notice that although the equations (168) and (169) for displacements and electric potential are uncoupled, the piezoelectric boundary value problem is coupled through the moment traction as indicated in (170).

## 6.2. Governing equations for two dimensional isotropic material

We suppose that the media occupies a cylindrical region, such that the axis of the cylinder is parallel to the $x_3$-axis. Furthermore, we assume the body is in a state of planar deformation and polarization parallel to this plane, such that

$$u_{\alpha,3} = 0,\ \phi_{,3} = 0,\ u_3 = 0 \quad \text{in} \quad V \tag{171a-c}$$

where all Greek indices here, and throughout the remainder of the paper, will vary only over (1,2). Also, let $V^{(2)}$ and $S^{(2)}$ represent, respectively, the cross section of the body in the $x_1 x_2$-plane and its bounding edge in that plane. Interestingly, if the location on the boundary contour in the *x-y* plane is specified by the coordinate *s* in a positive sense, we have $n_1 = \dfrac{dx_2}{ds}$ and $n_2 = \dfrac{dx_2}{ds}$. These can be written in the index form

$$n_\alpha = \varepsilon_{\alpha\beta} \frac{dx_\beta}{ds} \tag{172}$$

where $\varepsilon_{\alpha\beta}$ is the two-dimensional alternating or permutation symbol with

$$\varepsilon_{12} = -\varepsilon_{21} = 1, \quad \varepsilon_{11} = \varepsilon_{22} = 0 \tag{173}$$

As a result of these assumptions, all quantities are independent of $x_3$. Then, throughout the domain

$$\omega_\alpha = 0, \ e_{3i} = e_{i3} = 0, \ \kappa_3 = 0, \ E_3 = 0 \tag{174a-d}$$

and

$$\sigma_{3\alpha} = \sigma_{\alpha 3} = 0, \ \mu_3 = \mu_{21} = 0, \ D_3 = 0 \tag{175a-c}$$

Introducing the abridged notation, one can define

$$\omega = \omega_3 = \frac{1}{2}(u_{2,1} - u_{1,2}) = \frac{1}{2}\varepsilon_{\alpha\beta} u_{\beta,\alpha} \tag{176}$$

and the non-zero components of the curvature vector are

$$\kappa_\alpha = \frac{1}{2}\varepsilon_{\alpha\beta}\omega_{,\beta} \tag{177}$$

Therefore, the non-zero components of stresses and polarizations are written

$$\mu_\alpha = -8\mu l^2 \kappa_\alpha + 2fE_\alpha \tag{178}$$

$$D_\alpha = \epsilon E_\alpha + 4f\kappa_\alpha \tag{179}$$

$$\sigma_{\beta\alpha} = \lambda e_{\gamma\gamma}\delta_{\alpha\beta} + 2\mu e_{\alpha\beta} + 2\mu l^2 \varepsilon_{\alpha\beta}\nabla^2\omega \tag{180}$$

The components of the force-stress tensor in explicit form are

$$\sigma_{11} = \frac{2\mu}{1-2\nu}\left[(1-\nu)e_{11} + \nu e_{22}\right] \tag{181a}$$

$$\sigma_{22} = \frac{2\mu}{1-2\nu}\left[\nu e_{11} + (1-\nu)e_{22}\right] \tag{181b}$$

$$\sigma_{12} = 2\mu e_{12} - 2\mu l^2 \nabla^2 \omega \tag{181c}$$

$$\sigma_{21} = 2\mu e_{12} + 2\mu l^2 \nabla^2 \omega \tag{181d}$$

All the other components are zero, apart from $\sigma_{33}$ and $\mu_{3\alpha}$, which are given as

$$\sigma_{33} = \nu\sigma_{\gamma\gamma} = \nu(\sigma_{11} + \sigma_{22}) \tag{182}$$

$$\mu_{3\alpha} = -4\mu l^2 \omega_{,\alpha} \tag{183}$$

Notice that the stresses in (182) and (183) act on planes parallel to the $x_1 x_2$-plane.

For the planar problem, the stresses and polarizations must satisfy the following equations

$$\sigma_{\beta\alpha,\beta} + F_\alpha = \rho\ddot{u}_\alpha \tag{184}$$

$$\sigma_{[\beta\alpha]} = -\mu_{[\alpha,\beta]} \tag{185}$$

$$D_{\alpha,\alpha} = \rho_E \tag{186}$$

with the obvious requirement $F_3 = 0$. It should be noticed that the moment equation has given the non-zero components

$$\sigma_{[21]} = -\sigma_{[12]} = -\mu_{[1,2]} = 2\mu l^2 \nabla^2 \omega \tag{187}$$

in (181c,d).

Therefore, for the governing equations, we have

$$\left[\lambda + \mu(1 + l^2 \nabla^2)\right] u_{\beta,\beta\alpha} + \mu(1 - l^2 \nabla^2)\nabla^2 u_\alpha + F_\alpha = \rho\ddot{u}_\alpha \tag{188}$$

$$\epsilon\nabla^2 \phi + \rho_E = 0 \tag{189}$$

The force-traction reduces to

$$t_\alpha^{(n)} = \sigma_{\beta\alpha} n_\beta \tag{190}$$

and the moment-traction has only one component $m_3$. This can be conveniently denoted by the abridged symbol $m$, where

$$m = m_3^{(n)} = \varepsilon_{\beta\alpha}\mu_\alpha n_\beta = 4\mu l^2 \frac{\partial \omega}{\partial n} - 2f \frac{\partial \varphi}{\partial s} \tag{191}$$

For normal electric displacement, we have

$$d = -\epsilon \frac{\partial \varphi}{\partial n} + 2f \frac{\partial \omega}{\partial s} \tag{192}$$

## 6.3. Polarization of an isotropic dielectric cylinder

Consider a long isotropic homogeneous linear flexoelectric dielectric cylinder of radius $a$ placed perpendicular to an initially uniform electric field with magnitude $E_0$ in the $x_1$ direction, as indicated in Fig. 1. The polarization of the cylinder disturbs the electric field, such that the electric potentials for inside and outside cylinder in polar coordinates are

$$\phi = \begin{cases} \phi_{in} = A_1 r \cos\theta & r \leq a \\ \phi_{out} = \left(-E_0 r + B_2 \frac{1}{r}\right)\cos\theta & r \geq a \end{cases} \tag{193}$$

where at distances far from the cylinder

$$\phi \to -E_0 x_1 = -E_0 r \cos\theta \qquad \text{as } r \to \infty \tag{194}$$

Notice that these solutions satisfy the Laplace's equation (189) without free electric charges. As a result of the flexoelectricity effect, the cylinder also experiences some deformation and internal stresses. The elastostatic components of the displacements satisfy the equation (188) by neglecting the body force and the inertial effects. These components in polar coordinates are

$$u_r = \left[(1-4\nu)C_1 r^2 + C_2 \frac{1}{r} I_1\left(\frac{r}{l}\right) - \frac{1}{2l}C_2\right]\cos\theta \tag{195a}$$

$$u_\theta = \left\{(5-4\nu)C_1 r^2 - C_2\left[\frac{1}{l}I_0\left(\frac{r}{l}\right) - \frac{1}{r}I_1\left(\frac{r}{l}\right)\right] + \frac{1}{2l}C_2\right\}\sin\theta \tag{195b}$$

where $I_n$ is the modified Bessel function of first and second kind of order $n$, respectively. The four constants $A_1$, $B_2$, $C_1$ and $C_2$ are to be determined from boundary conditions. It should be noticed that the last terms in the expressions (195a,b) represent a rigid body translation in the $x_1$ direction, which has been added to keep the center of the cylinder to remain at the origin.

For components of the electric field, strain, rotation and mean curvature in polar coordinates, we obtain

$$E_r = -\frac{\partial \phi}{\partial r} = \begin{cases} -A_1 \cos\theta & r \leq a \\ \left(E_0 + B_2 \frac{1}{r^2}\right)\cos\theta & r \geq a \end{cases} \quad (196a)$$

$$E_\theta = -\frac{\partial \phi}{r\partial \theta} = \begin{cases} A_1 \sin\theta & r \leq a \\ \left(-E_0 + B_2 \frac{1}{r^2}\right)\sin\theta & r \geq a \end{cases} \quad (196b)$$

$$e_{rr} = \frac{\partial u_r}{\partial r} = \left[2(1-4\nu)C_1 r + C_2\left[\frac{1}{lr}I_0\left(\frac{r}{l}\right) - \frac{2}{r^2}I_1\left(\frac{r}{l}\right)\right]\right]\cos\theta \quad (196c)$$

$$e_{\theta\theta} = \frac{1}{r}\frac{\partial u_\theta}{\partial \theta} + \frac{u_r}{r} = \left\{2(3-4\nu)C_1 r^2 - C_2\left[\frac{1}{lr}I_0\left(\frac{r}{l}\right) - \frac{2}{r}I_1\left(\frac{r}{l}\right)\right]\right\}\cos\theta \quad (196d)$$

$$e_{r\theta} = \frac{1}{2}\left(\frac{1}{r}\frac{\partial u_r}{\partial \theta} + \frac{\partial u_\theta}{\partial r} - \frac{u_\theta}{r}\right) = \left[2C_1 r - \frac{1}{2}C_2\left[\frac{1}{l^2}I_1\left(\frac{r}{l}\right) - \frac{2}{lr}I_0\left(\frac{r}{l}\right) + \frac{2}{r^2}I_1\left(\frac{r}{l}\right)\right]\right]\sin\theta \quad (196e)$$

$$\omega = \frac{1}{2}\left(\frac{\partial u_\theta}{\partial r} + \frac{u_\theta}{r} - \frac{1}{r}\frac{\partial u_r}{\partial \theta}\right) = \frac{1}{2}\left[16(1-\nu)C_1 r - C_2\frac{1}{l^2}I_1\left(\frac{r}{l}\right)\right]\sin\theta \quad (196f)$$

$$\kappa_{rz} = \kappa_\theta = -\frac{1}{2}\frac{\partial \omega}{\partial r} = -\frac{1}{4l^2}\left\{16(1-\nu)l^2 C_1 - C_2\left[\frac{1}{l}I_0\left(\frac{r}{l}\right) - \frac{1}{r}I_1\left(\frac{r}{l}\right)\right]\right\}\sin\theta \quad (196g)$$

$$-\kappa_{\theta z} = \kappa_r = \frac{1}{2}\frac{1}{r}\frac{\partial \omega}{\partial \theta} = \frac{1}{4l^2}\left[16(1-\nu)l^2 C_1 - C_2\frac{1}{r}I_1\left(\frac{r}{l}\right)\right]\cos\theta \quad (196h)$$

For the components of force- and couple-stresses, and electric displacements, we have

$$\sigma_{rr} = \frac{2\mu}{1-2\nu}\left[(1-\nu)e_{rr} + \nu e_{\theta\theta}\right] = 2\mu\left\{2C_1 r + C_2\left[\frac{1}{lr}I_0\left(\frac{r}{l}\right) - \frac{2}{r^2}I_1\left(\frac{r}{l}\right)\right]\right\}\cos\theta \quad (197a)$$

$$\sigma_{\theta\theta} = \frac{2\mu}{1-2\nu}\left[\nu e_{rr} + (1-\nu)e_{\theta\theta}\right] = 2\mu\left\{6C_1 r - C_2\left[\frac{1}{lr}I_0\left(\frac{r}{l}\right) - \frac{2}{r^2}I_1\left(\frac{r}{l}\right)\right]\right\}\cos\theta \quad (197b)$$

$$\sigma_{r\theta} = 2\mu e_{r\theta} - 2\mu l^2 \nabla^2 \omega = 2\mu\left\{2C_1 r + C_2\left[\frac{1}{lr}I_0\left(\frac{r}{l}\right) - \frac{2}{r^2}I_1\left(\frac{r}{l}\right)\right]\right\}\sin\theta \quad (197c)$$

$$\sigma_{\theta r} = 2\mu e_{r\theta} + 2\mu l^2 \nabla^2 \omega = 2\mu\left\{2C_1 r - C_2\left[\frac{1}{l^2}I_1\left(\frac{r}{l}\right) - \frac{1}{lr}I_0\left(\frac{r}{l}\right) + \frac{2}{r^2}I_1\left(\frac{r}{l}\right)\right]\right\}\sin\theta \quad (197d)$$

$$-\mu_{\theta z} = \mu_r = -2\mu\left[16(1-v)l^2 C_1 - C_2 \frac{1}{r}I_1\left(\frac{r}{l}\right)\right]\cos\theta - 2fA_1\cos\theta \tag{197e}$$

$$\mu_{rz} = \mu_\theta = 2\mu\left\{16(1-v)l^2 C_1 - C_2\left[\frac{1}{l}I_0\left(\frac{r}{l}\right) - \frac{1}{r}I_1\left(\frac{r}{l}\right)\right]\right\}\sin\theta + 2fA_1\sin\theta \tag{197f}$$

$$D_r = \epsilon E_r + 4f\kappa_r$$

$$= \begin{cases} -\epsilon A_1 \cos\theta + \dfrac{f}{l^2}\left[16(1-v)l^2 C_1 - C_2 \dfrac{1}{r}I_1\left(\dfrac{r}{l}\right)\right]\cos\theta & r \leq a \\ \epsilon_0\left(E_0 + B_2 \dfrac{1}{r^2}\right)\cos\theta & r \geq a \end{cases} \tag{197g}$$

$$D_\theta = \epsilon E_\theta + 4f\kappa_\theta$$

$$= \begin{cases} \epsilon A_1 \sin\theta - \dfrac{f}{l^2}\left\{16(1-v)l^2 C_1 - C_2\left[\dfrac{1}{l}I_0\left(\dfrac{r}{l}\right) - \dfrac{1}{r}I_1\left(\dfrac{r}{l}\right)\right]\right\}\sin\theta & r \leq a \\ \epsilon_0\left(-E_0 + B_2 \dfrac{1}{r^2}\right)\sin\theta & r \geq a \end{cases} \tag{197h}$$

The electrical boundary conditions at $r = a$ are $\phi_{in} = \phi_{out}$ and $D_r|_{in} = D_r|_{out}$, which enforce the continuity of the electric potential and the electric displacement. These give

$$A_1 a = -E_0 a + B_2 \frac{1}{a} \tag{198a}$$

$$-\epsilon A_1 + \frac{f}{l^2}\left[16(1-v)l^2 C_1 - C_2 \frac{1}{a}I_1\left(\frac{a}{l}\right)\right] = \epsilon_0\left(E_0 + B_2 \frac{1}{a^2}\right) \tag{198b}$$

The three mechanical boundary conditions at $r = a$ are $\sigma_{rr} = 0$, $\sigma_{r\theta} = 0$ and $\mu_{rz} = 0$, which represent the traction-free surface. Interestingly, these give only two independent required equations

$$2C_1 + C_2 \frac{1}{a^2}\left[\frac{1}{l}I_0\left(\frac{a}{l}\right) - \frac{2}{a}I_1\left(\frac{a}{l}\right)\right] = 0 \tag{199a}$$

$$16(1-v)l^2 C_1 - C_2\left[\frac{1}{l}I_0\left(\frac{a}{l}\right) - \frac{1}{a}I_1\left(\frac{a}{l}\right)\right] + \frac{f}{\mu}A_1 = 0 \tag{199b}$$

Using the relations (198) and (199), we can obtain the four unknown coefficients $A_1$, $B_2$, $C_1$ and $C_2$ to complete the solution.

For the numerical study, we select the following nondimensional values for the material parameters: $\mu = 1$, $\nu = 1/4$, $a = 1$, $l = 0.1$, $\epsilon_0 = 1$, $\epsilon = 2$, $f = 0.1$. Then,

$$A_1 = -0.6348 \tag{200a}$$

$$B_2 = 0.8892 \tag{200b}$$

$$C_1 = 0.0471 \tag{200c}$$

$$C_2 = -4.1275 \times 10^{-6} \tag{200d}$$

The analytical solutions along the radial line at $\theta = \pi/2$ for radial stress $\sigma_{rr}$, and along the radial line at $\theta = 0$ for shear stress $\sigma_{r\theta}$ and couple-stress $\mu_{zr}$ are displayed in Figs. 2, 3, and 4, respectively. Notice that the uniform electric field has generated polarization, force- and couple-stresses in the isotropic cylinder.

## 7. Conclusions

The consistent size-dependent continuum mechanics is a practical theory, which enables us to develop many different formulations that may govern the behavior of solid continua at the smallest scales. The size-dependent electromechanical formulations have the priority because of their importance in nanomechanics and nanotechnology. Here, we have developed the size-dependent piezoelectricity, which shows the possible coupling of polarization to the mean curvature tensor. The most general anisotropic linear elastic material is described by 78 independent constitutive coefficients. This includes nine flexoelectric coefficients relating mean curvatures to electric displacements, and electric field components to the couple-stresses.

In addition, we have developed the corresponding weak forms, energy potentials, uniqueness theorem, and reciprocal theorem for linear piezoelectricity. The new size dependent piezoelectricity clearly shows that the piezoelectric effect can exist in isotropic couple stress materials, where the two Lamé parameters, one length scale, and one flexoelectric parameter completely characterize the behavior. The details for the general two-dimensional isotropic case are also elucidated. Finally, we have examined the polarization of an isotropic long cylinder in a uniform electric field and obtained the closed form solution.

The present theory shows that couple-stresses are necessary for the development of any electromechanical size dependent effect. Additional aspects of linear piezoelectricity, including fundamental solutions and computational mechanics formulations, will be addressed in forthcoming work. Beyond this, the present theory should be useful for the development of other size-dependent electromechanical formulations, such as piezomagnetism and magnetostriction, which are also important for analysis at small scales.

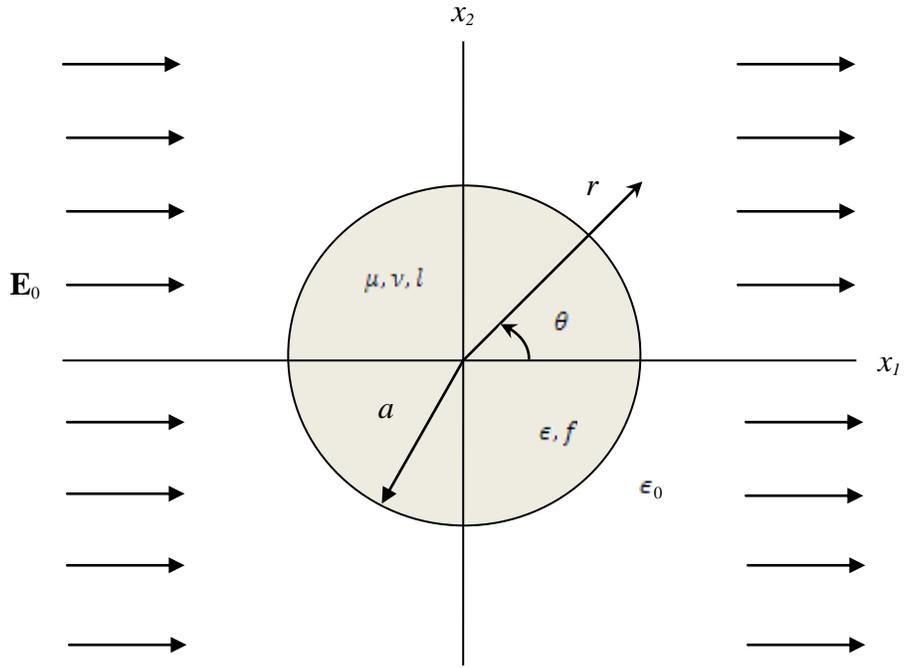

**Fig. 1.** Flexoelectric dielectric cylinder in uniform electric field.

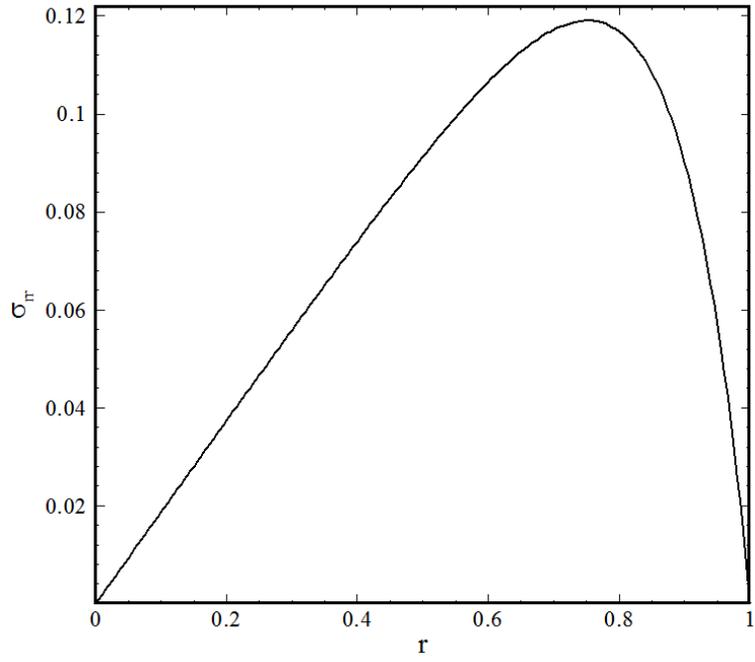

**Fig. 2.** Flexoelectric dielectric cylinder in uniform electric field. Radial stress $\sigma_{rr}$ on $\theta = \pi/2$.

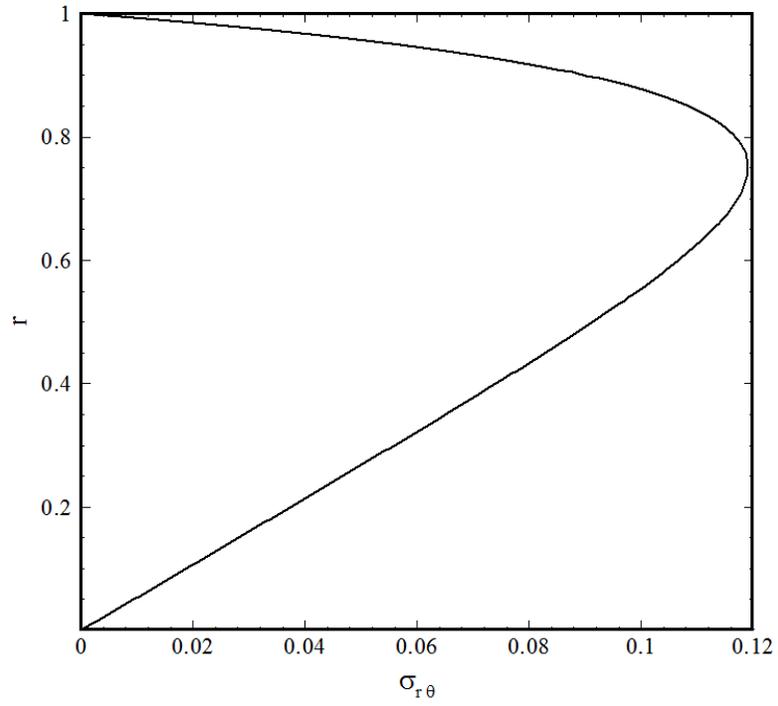

**Fig. 3.** Flexoelectric dielectric cylinder in uniform electric field. Shear stress $\sigma_{r\theta}$ on $\theta = 0$.

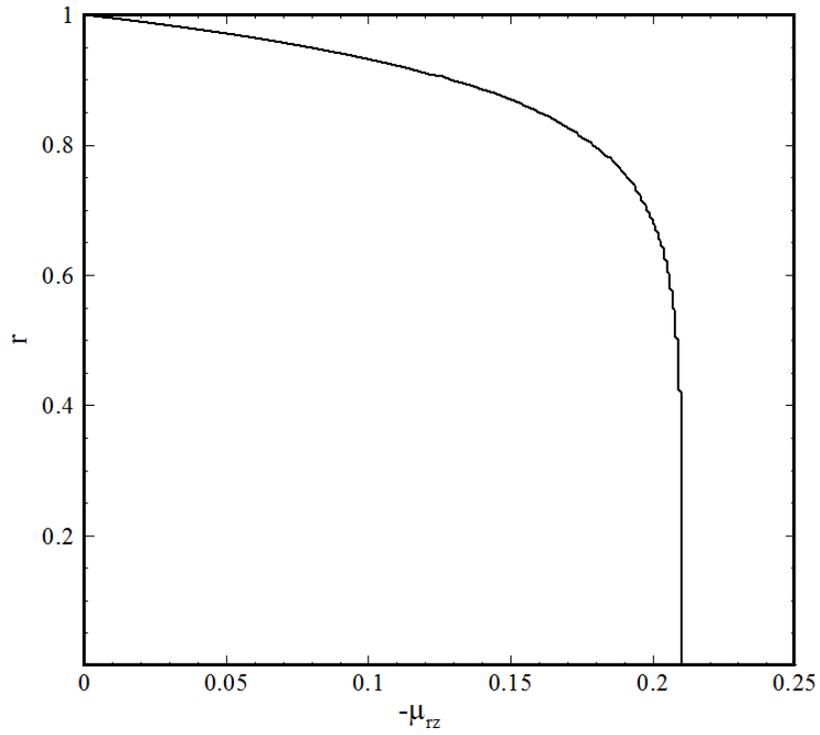

**Fig. 4.** Flexoelectric dielectric cylinder in uniform electric field. Couple-stress $\mu_{zr} = -\mu_{rz}$ on $\theta = 0$.